\documentclass[aps,prb,groupedaddress,onecolumn]{revtex4}
\usepackage{graphicx,wasysym}

\def\be{\begin{equation}}
\def\ee{\end{equation}}
\def\ba{\begin{eqnarray}}
\def\ea{\end{eqnarray}}
\def\bc{\begin{center}}
\def\ec{\end{center}}

\def\p{{\partial}}
 
\begin{document}

\title{Nonlinear broadening of the plasmon linewidth in a graphene stripe}

\author{S. A. Mikhailov and D. Beba}

\affiliation{Institute for Physics, University of Augsburg, D-86135 Augsburg, Germany}

\date{\today}

\begin{abstract}
In contrast to semiconductor structures, the experimentally observed plasma resonances in graphene show an asymmetrical and rather broad linewidth. We show that this can be explained by the linear electron energy dispersion in this material and is related to the violation of the generalized Kohn theorem in graphene.
\end{abstract}


\maketitle

\section{Introduction}

Plasma oscillations in semiconductor two-dimensional (2D) electron systems have been studied since late 1960-ies \cite{Stern67,Allen77,Theis78}. Plasmon related absorption resonances have been observed in different semicondutor structures (inversion layers\cite{Theis80}, quantum wells\cite{Heitmann86}, wires\cite{Demel88,Demel91,Goni91,Kukushkin05}, dots\cite{Allen83,Demel90,Heitmann92a,Kukushkin03a}, antidots\cite{Kern91,Heitmann92a,Hochgrafe99}, rings\cite{Dahl93,Kovalskii06}) and in the broad frequency range from microwaves (the frequency $f\simeq 10$ GHz) up to far-infrared ($f\simeq 1$ THz). Both the frequency and the linewidth of plasma resonances were found in good agreement with theoretical predictions.

In graphene\cite{Novoselov04,Novoselov05,Zhang05}, a new truly 2D material with very attractive physical properties\cite{Neto09}, the spectrum of plasma waves was first calculated in 2006-2007 \cite{Wunsch06,Hwang07}, see also \cite{Falkovsky07a,Falkovsky07b,Ryzhii07,Polini08,Hanson08,Rana08,Hill09,Popov10,Dubinov11,Ryzhii12}. However, the number of experimental studies of graphene plasmons is not very large, in spite of great interest to this material. Their observation by the electron energy loss \cite{Liu08,Langer10} and far-infrared transmission \cite{Ju11,Crassee12} spectroscopy has been reported only in few papers, see also \cite{Koppens11}. In all experimental studies the linewidth of plasmon resonances was found to be comparable with or even larger than their frequency, although graphene samples with the mobility of 200000 cm$^2$/Vs are currently available. 

Recently, it was theoretically shown \cite{Mikhailov09b} that the width of the classical cyclotron resonance (CR) should be substantially broader in graphene than in semiconductors due to the strongly non-parabolic (linear) energy dispersion of graphene electrons
\be 
{\cal E}_{\bf p}=\pm v_F|{\bf p}| \label{lin-spec}
\ee
(here $v_F\approx 10^8$ cm/s is the Fermi velocity). In fact, it was found that the spectrum (\ref{lin-spec}) leads to a broad observed CR linewidth even in perfectly pure graphene, in the absence of any scattering or other energy-loss mechanisms. The goal of this paper is to study whether the linear electronic spectrum (\ref{lin-spec}) may be also responsible for the broadening of plasma resonances in graphene. Using a non-perturbative kinetic approach we study the motion of massless quasiparticles (\ref{lin-spec}) in a finite-width stripe of graphene with a parabolic confinement potential and show that, indeed, the observed plasmon resonance line should be strongly asymmetric and have the width comparable with the plasmon frequency. This qualitatively agrees with observations of Refs. \cite{Liu08,Langer10,Ju11,Crassee12}.

\section{Plasma oscillations in a graphene stripe}

\begin{figure}
\includegraphics[width=8.5cm]{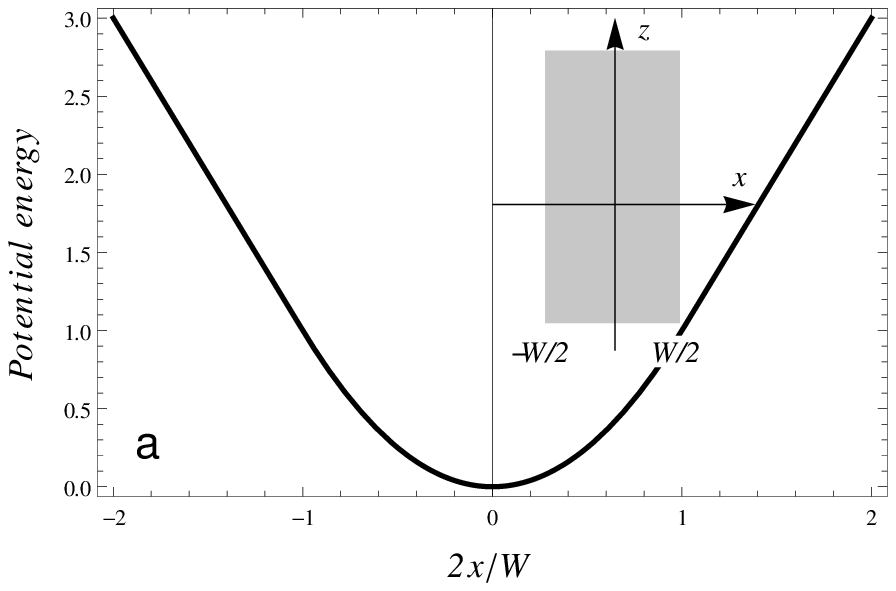}\\
\includegraphics[width=8.5cm]{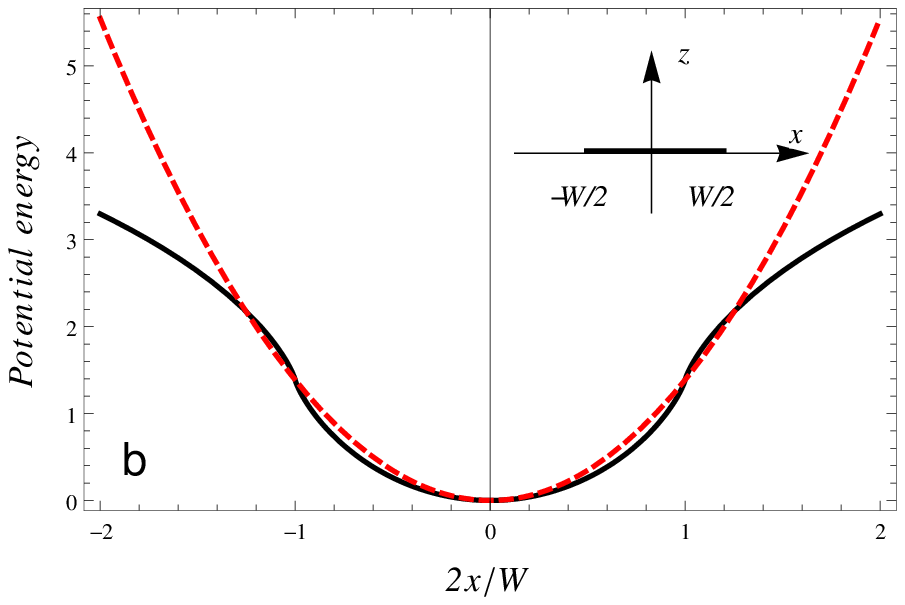}
\caption{\label{geom} The confining potential energy $-e\phi(x)$ due to the positive background in (a) a 3D slab and (b) a 2D stripe. Insets illustrate the geometry. Red dashed curve in (b) shows the parabolic approximation (\ref{Kapprox}).
}
\end{figure}

The plasma oscillations in a solid can be modeled as a collective motion of electrons in an external potential produced by a positively charged background. For example, in a three-dimensional metallic slab occupying the area $|x|<W/2$, see inset to Figure \ref{geom}a, the positive background with the charge density $+en_v\theta(W/2-|x|)$ creates the electric potential 
\be 
\phi(x)=-\frac{\pi en_vW^2}2\left\{
\begin{array}{ll}
(2x/W)^2,\ \ \ & \ \ \ |2x/W|\le 1,\\
4|x|/W-1, \ \ \ & \ \ \ |2x/W|>1.\\
\end{array}
\right. ,
\label{potential-slab}
\ee
shown in Figure \ref{geom}a. Being shifted from their equilibrium positions in the $x$-direction, electrons of the slab move in the potential (\ref{potential-slab}) in accordance with the equation of motion 
\be 
m\ddot x=e\frac{\p \phi}{\p x}=-4\pi n_ve^2 x, \ \ \ |x|<W/2.
\label{mx..}
\ee
Since the potential inside the slab is parabolic, Eq. (\ref{mx..}) describes harmonic oscillations with the frequency $\omega_{p3}^2=4\pi n_ve^2/m$ which coincides with the frequency of three-dimensional plasmons. 

In a 2D stripe $|x|<W/2$, $z=0$ (inset to Figure \ref{geom}b) the positive background with the charge density $+en_s\theta(W/2-|x|)\delta(z)$ creates the electric potential 
\be 
\phi(x,z=0)= -en_sW\left[\left(1-\frac {2x}W\right)\ln\left|1-\frac {2x}W\right|+\left(1+\frac {2x}W\right)\ln\left|1+\frac {2x}W\right| \right]
\label{pot2D}
\ee
which can be also approximated, with a good accuracy, by a parabolic form inside the stripe,
\be 
-e\phi(x,z=0)\approx Kx^2/2, \ \ \ |x|\le W/2,
\label{Kapprox}
\ee
Figure \ref{geom}b. Choosing the potential curvature $K$ from the condition 
$$
\int_0^{W/2}\left(-e\frac{\p \phi(x,z=0)}{\p x}- Kx\right)dx=0,
$$
so that the approximation (\ref{Kapprox}) creates the same average force inside the stripe, one gets 
\be 
K=16\ln 2\frac{n_se^2}W.
\label{Kvalue}
\ee 
The motion of all electrons in the stripe is then described by the harmonic oscillator equation of the type (\ref{mx..}) with the squared oscillation frequency $\omega_{p2}^2= K/m=16\ln 2n_se^2/mW$, which differs by only 10\% from the numerically calculated \cite{Mikhailov05a} plasma frequency in a finite-width 2D stripe. 

Due to the linear energy dispersion (\ref{lin-spec}) the dynamics of graphene quasi-particles substantially differs from that of electrons in semiconductors. In the absence of scattering their oscillations in the field of the positive background can be described by equations 
\be
\dot {\bf r}  =  v_F {\bf p}/p, \ \ \ \dot {\bf p} = {\bf F}=(-Kx){\bf e}_x, 
\label{eqgr}
\ee
with the curvature $K$ given by (\ref{Kvalue}). The problem (\ref{eqgr}) should be solved with initial conditions ${\bf r}(0)={\bf r}_0$ and ${\bf p}(0)={\bf p}_{0}=p_{0}(\cos\phi_0,\sin\phi_0)$, where $p_0\le p_F$, i.e. we assume that the momenta of all particles at $t=0$ lie inside the Fermi circle.

\begin{figure}
\includegraphics[width=8.5cm]{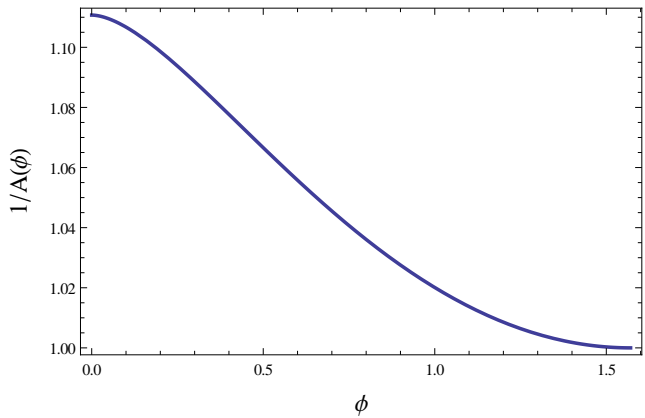}
\caption{\label{intA} The function $1/A(\phi)$ in the interval from $\phi=0$ to $\phi=\pi/2$. 
}
\end{figure}

The system of equations (\ref{eqgr}) has two integrals of motion, the momentum $p_{y0}$ and the total energy 
\be 
{\cal H}_0=v_F\sqrt{p_{x}^2+p_{y}^2}+K\frac {x^2}2=v_Fp_{0}+K\frac {x_0^2}2.
\label{1stIntEnergy}
\ee
The time dependence of the coordinate $x(t)$ is then determined by the formula 
\be 
t=\pm\int_{x_0}^{x}\frac {{\cal H}_0-Kx'^2/2}{v_F\sqrt{({\cal H}_0-Kx'^2/2)^2-v_F^2p_{y0}^2}}dx' ,
\label{solx}
\ee
the relation $p_x(t)$ can be found from (\ref{1stIntEnergy}), and the dependence $y(t)$ -- from the equation for $\dot y$ in (\ref{eqgr}). Equation (\ref{solx}) describes oscillations of the $x$-coordinate of the particle between two turning points $x_{1,2}=\pm\sqrt{2({\cal H}_0-v_F|p_{y0}|)/K}$. The oscillation period $T=T( p_{y0}, {\cal H}_0) =T(p_0,\phi_0,x_0)$ is then given by the formula
\be 
T =2 \int_{x_1}^{x_2}\frac {{\cal H}_0-Kx^2/2}{v_F\sqrt{({\cal H}_0-Kx^2/2)^2-v_F^2p_{y0}^2}}dx.
\ee
Taking the limit $x_0\to 0$ we finally get the oscillation frequency $\Omega=2\pi/T$ in the regime of very small oscillation amplitudes:
\be 
\Omega(p_{0}, \phi_0) =
\sqrt{\frac{ v_FK}{p_0}}\frac 1{A(\phi_0)},
\label{osc-freq}
\ee
where the function 
\be 
A(\phi)=
\frac{2\sqrt{2}}{ \pi}\int_{0}^{\sqrt{1-|\sin\phi|}}
\frac {1-x^2}{\sqrt{(1-x^2)^2-\sin^2\phi}}dx
\label{funcA}
\ee
has the properties $A(\phi)=A(-\phi)=A(\pi+\phi)$, $A(0)=2\sqrt{2}/\pi\approx 0.900$, $A(\pi/2)=1$. 
The inverse function $1/A(\phi)$ which determines the frequency (\ref{osc-freq}) is shown in Figure \ref{intA}.

\begin{figure}
\includegraphics[width=8.5cm]{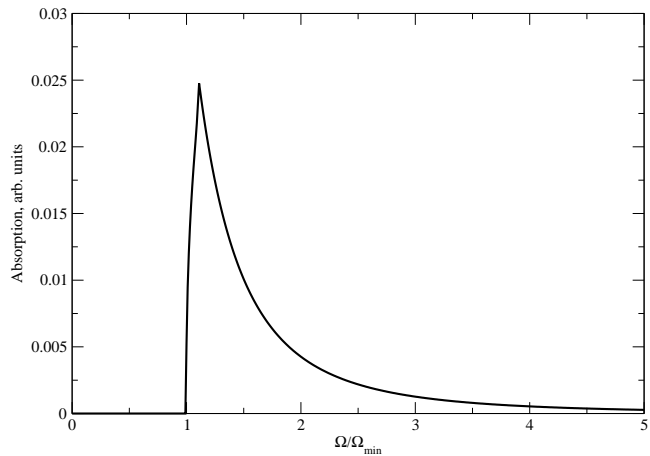}
\caption{\label{linewidth} The shape of the plasmon absorption line in a perfectly clean graphene stripe. 
}
\end{figure}

In contrast to semiconductors, the oscillation frequency $\Omega$ of massless quasiparticles, Eq. (\ref{osc-freq}), depends on the initial conditions. It weakly varies with $\phi_0$ (Figure \ref{intA}) but quite essentially depends on the absolute value of the initial momentum $p_0$. Particles with the largest (Fermi) momentum have the lowest oscillation frequency 
\be 
\Omega_{min} =
\sqrt{\frac{ v_FK}{p_F}} = 4\sqrt{\ln 2\frac{ n_se^2v_F^2}{E_FW}},
\label{freq-min}
\ee
see Figure \ref{linewidth}. Particles with smaller momenta oscillate with higher frequencies, therefore the observed 2D plasmon resonance line should have a broad asymmetric form with a flatter high-frequency and a steeper low-frequency side, Figure \ref{linewidth}. The maximum of the absorption plasmon curve lies at the frequency 
\be 
\Omega_p=\frac \pi{2\sqrt{2}}\Omega_{min} =
\sqrt{2\pi^2\ln 2\frac{ n_se^2v_F^2}{E_FW}}
\label{PlasFreqGraph}
\ee
which can be treated as the 2D plasmon frequency in a graphene stripe. The formula (\ref{PlasFreqGraph}) gives correct dependencies on all physical parameters and a reasonable numerical value for the plasma frequency in a graphene stripe (cf. with \cite{Mikhailov05a}). 

The shape of the plasmon absorption line (Figure \ref{linewidth}) is very similar to the one that was experimentally observed in Refs. \cite{Ju11,Crassee12}. Similar to the classical cyclotron resonance \cite{Mikhailov09b}, the 2D plasmon line broadening takes place in the collisionless limit, in the absence of any scattering in the graphene sample. Taking into account the finite scattering rate would lead to a somewhat smoother curve and to an additional broadening of the line. The collisionless broadening of the 2D plasmon resonance line is a direct consequence of the linear energy dispersion (\ref{lin-spec}) and is related with the violation of the generalized Kohn theorem \cite{Kohn61,Brey89,Maksym90,Govorov91,Peeters90,Mikhailov98a} in electronic systems with a non-parabolic energy dispersion. 

\section{Discussion and summary}

Our results for the {\em frequency} of the 2D plasmon resonance (\ref{PlasFreqGraph}) are in very good agreement with those of Refs. \cite{Wunsch06,Hwang07}. As for the {\em linewidth} of the resonance, they are substantially different: our calculations give a finite-width resonance line in the collisionless limit, while those of Refs. \cite{Wunsch06,Hwang07} give a vanishing linewidth under the same conditions. We believe that the reason of this difference results from the non-analyticity of the graphene spectrum (\ref{lin-spec}) in the Dirac point ${\bf p=0}$. The solution of the equations of motion (\ref{eqgr}) is equivalent to a {\em non-perturbative} solution of the Boltzmann kinetic equation. In Refs. \cite{Wunsch06,Hwang07} the {\em linear response} approach was applied, i.e. the solution of the kinetic equation was expanded, from the very beginning, in powers of the electric field and only the first-order term was taken into account. Mathematically, such a procedure suggests that all coefficients of the kinetic equation are analytical functions of ${\bf p}$, which is the case in semiconductor structures but is not the case in graphene. In order to fully understand the graphene response to external electromagnetic fields a more detailed  {\em non-perturbative} classical (Boltzmann) or quantum theory is thus required. 

To summarize, we have predicted that the 2D plasmon resonance line should be asymmetric and substantially broadened, even in perfectly clean graphene (without disorder) due to the linear energy dispersion of graphene electrons. This broadening is not the case in semiconductor systems with a parabolic energy dispersion and is related to the violation of the generalized Kohn theorem in graphene. 

The work was supported by Deutsche Forschungsgemeinschaft.


\end{document}